\documentclass{webofc}
\usepackage[varg]{txfonts}   
   \usepackage{grffile} 
    \usepackage{color}
   
   \newcommand{\arxivold}[1][]{[#1]}
   \newcommand{\arXiv}[1][]{[#1]}
   \newcommand{\arXivold}[1][]{[#1]}

\newcommand*{\refdoi}[1]{
  \href{https://doi.org/#1}{doi:#1}
}

\usepackage{xspace}                    
\xspaceaddexceptions{[]}
\usepackage{amssymb}
\usepackage{amsmath}
\usepackage{pbox}
\usepackage{bm}                        

\newcommand{\disc}{\discretionary{}{}{}}
\newcommand{\absatz}{\vspace{2ex}\noindent}

\newcommand{\ie}{\textit{i.e.}\xspace}
\newcommand{\etal}{\textit{et al.}\xspace}

\newcommand{\hqqq}{\hspace{2em}}

\newcommand{\vectorwithspace}[1]{\vec{#1}\mkern2mu\vphantom{#1}}

\newcommand{\pv}{\vectorwithspace{p}}
\newcommand{\qv}{\vectorwithspace{q}}

\newcommand{\mpi}{\ensuremath{m_\pi}}

\newcommand{\MeV}{\ensuremath{\mathrm{MeV}}}

\newcommand{\ChiEFT}{$\chi$EFT\xspace}

\newcommand{\NXLO}[1]{N\ensuremath{{}^{#1}}LO\xspace}

\newcommand{\HIGS}{HI$\gamma$S\xspace}
\newcommand{\threeH}{\ensuremath{{}^3}H\xspace}
\newcommand{\threeHe}{\ensuremath{{}^3}He\xspace}
\newcommand{\fourHe}{\ensuremath{{}^4}He\xspace}

\newcommand{\alphae}{\ensuremath{\alpha_{E1}}}
\newcommand{\betam}{\ensuremath{\beta_{M1}}}

\newcommand{\alphaep}{\ensuremath{\alpha_{E1}^{(\mathrm{p})}}}

\newcommand{\alphaen}{\ensuremath{\alpha_{E1}^{(\mathrm{n})}}}

\newcommand{\alphaes}{\ensuremath{\alpha_{E1}^{(\mathrm{s})}}}
\newcommand{\betams}{\ensuremath{\beta_{M1}^{(\mathrm{s})}}}

\newcommand{\omegalab}{\ensuremath{\omega_\mathrm{lab}}}

\newcommand{\thetalab}{\ensuremath{\theta_\mathrm{lab}}}

\newcommand{\chiSMSfourfive}{$\chi$SMSN$^4$LO+$450\MeV$+N$^2$LO3NI\xspace}

\newcommand{\calO}{\mathcal{O}}

\begin{document}
\title{Using the Transition-Density Formalism in the \\First Computation of 4He
  Compton Scattering}
%
%

\author{\firstname{Harald W.}
  \lastname{~Grie\3hammer}\inst{1}\fnsep\thanks{\email{hgrie@gwu.edu}; presenter}
  \and
        \firstname{Junjie} \lastname{Liao}\inst{1}\and
        \firstname{Judith A.} \lastname{McGovern}\inst{2} \and
        \firstname{Andreas} \lastname{Nogga}\inst{3} \and
        \firstname{Daniel R.} \lastname{Phillips}\inst{4}
}

\institute{Institute for Nuclear Studies, Department of Physics, The
    George Washington University, Washington DC 20052, USA
\and
            School of Physics and Astronomy, The University of
    Manchester, Manchester M13 9PL,  UK
\and
            IAS-4, IKP-3 and JCHP, Forschungszentrum J\"ulich, D-52428
    J\"ulich, Germany
\and
            Department of Physics and Astronomy and Institute of Nuclear
    and Particle Physics, Ohio University, Athens OH 45701, USA
          }

          \abstract{%
            The method and results of the first theory description of \fourHe
            Compton scattering at nuclear energies is presented, with a focus
            on figures. An upcoming publication~\cite{ourstuff} contains
            details and a comprehensive list of references. 
          }
\maketitle
%

Compton scattering on light nuclei 
up to about the first resonance region serves a dual purpose: to extract the
neutron polarisabilities; and to explore the part of the nuclear interactions
mediated by charged pion-exchange currents. Indeed, these two are interwoven:
Neutron properties are more easily extracted when bound (and therefore stable
inside the nucleus), but only if the nuclear binding effects are accounted for
by model-independent and accurate theory. In Compton scattering from $50$ to
about $130\;\MeV$, the nucleon's stiffness against deformation in electric and
magnetic fields is measured by the parameters of its two-photon response: the
electric and magnetic dipole polarisabilities $\alphae$ and $\betam$ reveal
the extent to which charge and current distributions of the nucleon's
constituents shift in external electromagnetic fields, characterising the
induced radiation dipoles~\cite{Griesshammer:2012we}. The very same
constituents are also responsible for one major part of the interactions which
bind nucleons into nuclei.

Chiral Effective Field Theory (\ChiEFT) is the model-independent and
systematically improvable framework in which such questions can be
systematically answered from data, namely with credible estimates of residual
theory uncertainties. Indeed, a large-scale international effort at a new
generation of high-precision facilities aims to understand low-energy Nuclear
Physics by extracting nucleon polarisabilities from Compton scattering
experiments, in close collaboration between theory and experiment; see also
the contributions by G.~Feldman~\cite{Feldman} and D.~Hornidge~\cite{Hornidge}
reporting on the \HIGS and MAMI A2 parts, respectively. So far, the isoscalar
(isospin-averaged) polarisabilities are found from deuteron
data~\cite{Myers:2014ace,Myers:2015aba}:
\begin{equation}
\label{eq:alphabeta}
  \alphaes=11.1\pm0.6_\mathrm{stat}\pm0.2_\mathrm{B\Sigma R}\pm0.8_\mathrm{th}\;\;,\;\;
\betams=\phantom{0}3.4\mp0.6_\mathrm{stat}\pm0.2_\mathrm{B\Sigma R}\pm0.8_\mathrm{th}\;\;.
\end{equation}
To reduce the combined uncertainties so that they become commensurate to those
of the proton polarisabilities of $\pm0.5$, both high-quality data and
high-accuracy theory are needed. 
This will provide a better understanding of the degree to which degree proton
and neutron polarisabilities differ and check a prediction based on a
Cottingham Sum Rule evaluation of the self-energy correction to the
proton-neutron mass difference:
$\alphaep-\alphaen=[-1.7\pm0.4]$~\cite{Gasser:2015dwa}.

Different nuclear targets test different linear combinations of proton and
neutron polarisabilities. For example, the deuteron and \fourHe are sensitive
to the isospin averages, while \threeHe probes $2\alphaep+\alphaen$ etc,
allowing for cross-validation of high-accuracy measurements and
theory. Furthermore, rates increase with atomic number. The deuteron theory is
well-understood~\cite{Griesshammer:2012we}, and \threeHe computations are now
available as well~\cite{Choudhury:2007bh, Shukla:2008zc, ShuklaPhD,
  Margaryan:2018opu}. On top of available deuteron, \fourHe and $^6$Li data,
new experiments are approved for these targets and \threeHe~\cite{Feldman,
  Hornidge}.

As a target, \fourHe has several endearing features. For experimentalists, it is
cheap, inert, safe to handle, liquefies at relatively high temperatures, and
its high dissociation energy allows clear differentiation between
elastic and inelastic events even with detectors of modest energy
resolution. For theorists, this perfect scalar-isoscalar target
allows extractions of the scalar-isoscalar polarisabilities which are free of
contamination from spin polarisabilities, and for a high-accuracy test of
the charged-meson exchange currents in a tightly bound system.

\begin{figure}[!b]
\centering
\includegraphics[width=\linewidth,clip]
{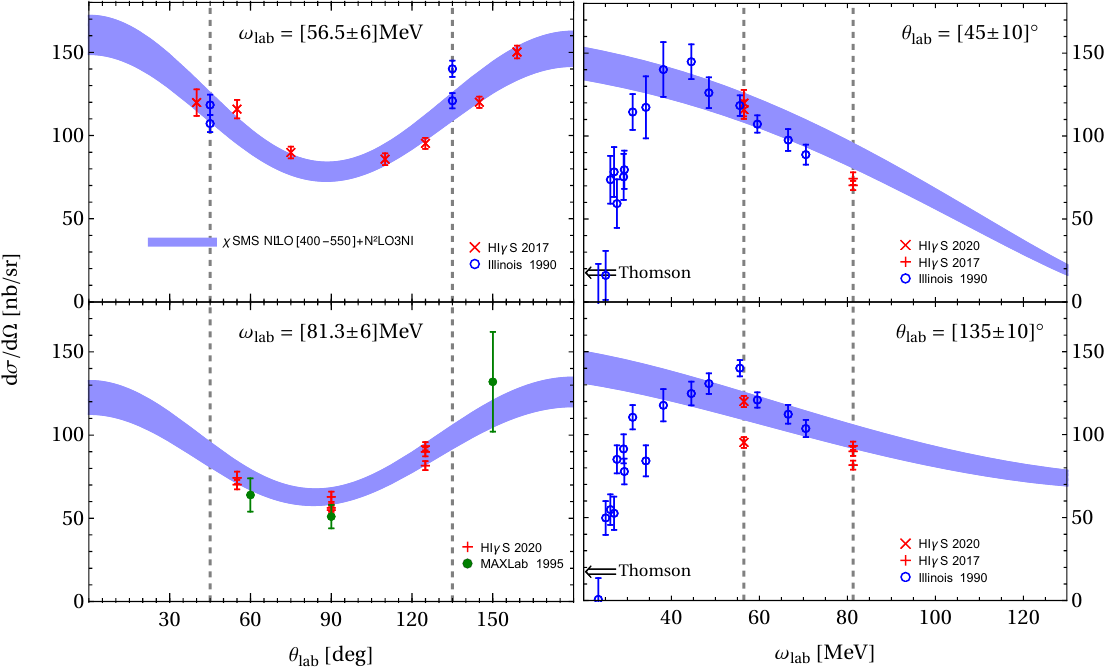}
\caption{(Colour on-line) \fourHe Compton cross sections at
  $\calO(e^2\delta^3)$ [\NXLO{3}] at $\omegalab=56.5\;\MeV$ (top left) and
  $81.3\;\MeV$ (bottom left), and $\thetalab=45^\circ$ (top right) and
  $135^\circ$ (bottom right)~\cite{ourstuff}. The band is generated with
  nucleon densities from $\chi$SMS\NXLO{4}+N$^2$LO3NI potentials with cutoffs
  $\{400;450;500;550\}\;\MeV$~\cite{Reinert:2017usi}. Data from Illinois~\cite{WellsPhD},
  MAXlab~\cite{Fuhrberg:1995zz} and \HIGS~\cite{Sikora:2017rfk, Li:2019irp}
  are included without accounting for differences between nominal and actual
  energies or angles. Central polarisability values as in
  eq.~\eqref{eq:alphabeta}. Lack of agreement $\lesssim50\;\MeV$ expected;  see ref.~\cite{ourstuff} for
  details. 
}
\label{fig:data}       
\end{figure}

\absatz Figure~\ref{fig:data} summarises the key result%
, with the isoscalar polarisabilities of eq.~\eqref{eq:alphabeta} as
input~\cite{ourstuff}. The blue band produced by chiral ``semilocal
momentum-space regularised'' (SMS) potentials at different
cutoffs~\cite{Reinert:2017usi} suggests that a \ChiEFT treatment at order
$\calO(e^2 \delta^3)$ [\NXLO{3}] has less-than-$10\%$ uncertainties from
residual dependence on the details of chiral $2$N and $3$N interactions.
Order-by-order convergence indicates an even smaller lower-bound uncertainty
of about $\pm6\%$ at the highest energies. Agreement between theory and the
$50$ available data from \HIGS, MAXlab and the University of Illinois Tagged
Photon Facility~\cite{Fuhrberg:1995zz,WellsPhD,Sikora:2017rfk,Li:2019irp} is
good within the experimental and theoretical uncertainties \emph{in the range
  where the assumptions of the present theoretical description hold}: namely
for those $\omega\sim\mpi$ at which the intermediate four-nucleon system
predominantly propagates incoherently, with only minor rescattering effects.
These only become dominant as $\omega\to0$ to restore the Thomson
limit. Informed by these data and accounting for uncertainties of both theory
and experiment, it is safe to conclude that the incoherent-propagation
assumption is justified at $\omega\gtrsim50\;\MeV$. A thorough discussion of
uncertainties and convergence of the \ChiEFT expansion is left to an upcoming
publication~\cite{ourstuff}.

\begin{figure}[!t]
\centering
\includegraphics[width=\linewidth,clip]
{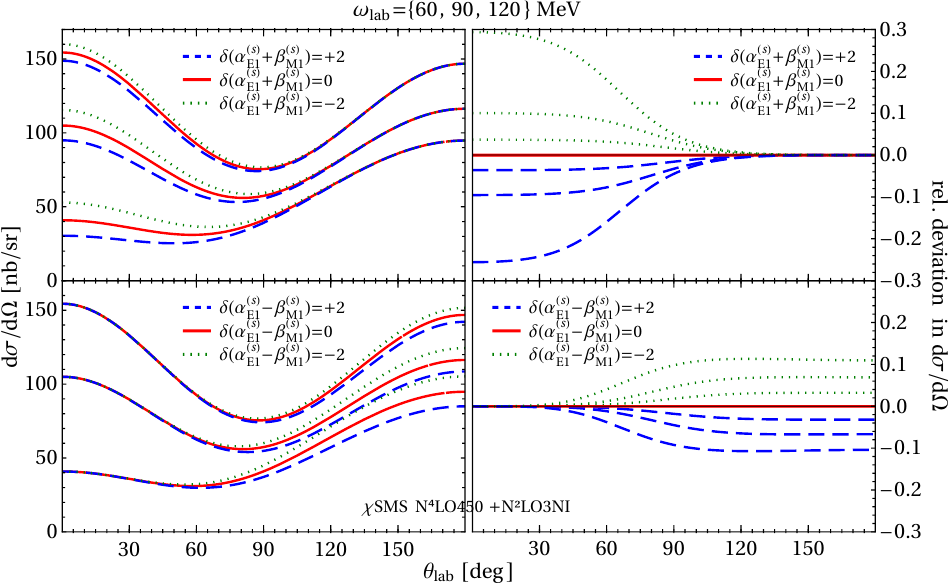}
\caption{(Colour on-line) Sensitivity of the cross section to varying the
  scalar-isoscalar polarisabilities around their central values (solid line)
  of eq.~\eqref{eq:alphabeta} by $+2$ (blue dashed) and $-2$ (green doted)
  units for a ``mean'' potential \chiSMSfourfive at $60$, $90$ and $120\;\MeV$
  (top to bottom)~\cite{ourstuff}. Left: differential cross section. Right:
  Relative deviations from central values increase with energy.}
\label{fig:polsvar}       
\end{figure}

For planning experiments, theory uncertainties are mitigated because they are
mostly angle-independent, whereas fig.~\ref{fig:polsvar} demonstrates that the
absolute and relative \emph{sensitivities} to varying the scalar-isoscalar
polarisability combinations $\alphaes\pm\betams$ have a rather strong angular
dependence: $\alphaes+\betams=14.5\pm0.4$ is fairly well constrained by the
Baldin Sum Rule, while the uncertainties in $\alphaes-\betams$ dominate
eq.~\eqref{eq:alphabeta}. Thus, these results are sufficiently
reliable to be useful for an exploratory study of magnitudes and sensitivities
to the nucleon polarisabilities to advance current planning of experiments --
as previously argued for \threeHe~\cite{Margaryan:2018opu}. Polarisability
extractions, on the other hand, should address residual theoretical
uncertainties more diligently, as for the proton and
deuteron~\cite{Griesshammer:2015ahu}. That work is under way, most notably to
include rescattering effects~\cite{Walet:2023myk} and update the Compton
kernels to $\calO(e^2\delta^4)$ [\NXLO{4}].

\begin{figure}[!t]
\centering
\includegraphics[width=0.86\linewidth,clip]
{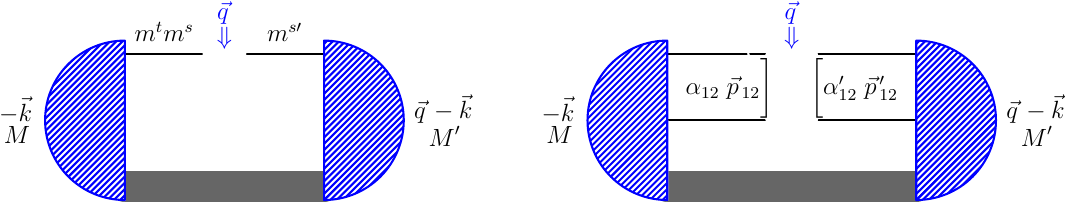}
\\[2ex]
\includegraphics[width=0.68\linewidth,clip]
{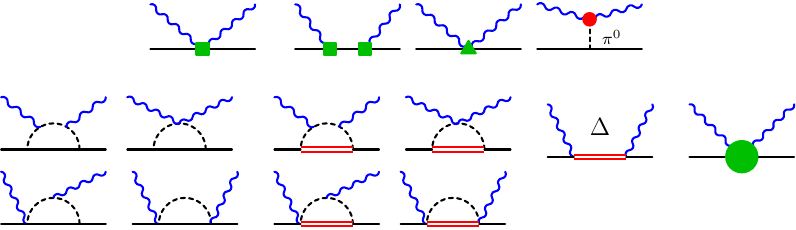}\hqqq
\includegraphics[width=0.22\linewidth,clip]
{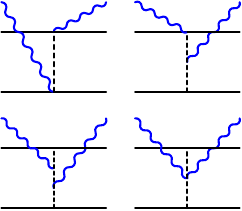}
\caption{(Colour on-line) Top: Definition of onebody (left) and twobody
  (right) densities in the $A$-nucleon bound state in the cm
  system~\cite{ourstuff}. One-nucleon quantum numbers are the spin and isospin
  projections $m^s$ and $m^t$ of the active nucleon. The two-nucleon quantum
  numbers are collectively denoted as $\alpha_{12}$ and relative momentum
  $\pv_{12}$. The spin projection of the incident nucleus is $M$. Primed
  variables for outgoing states. Bottom: Onebody (left) and twobody (right)
  Compton kernels up to and including order $\calO(e^2\delta^3)$
  [\NXLO{3}]~\cite{Choudhury:2007bh, Shukla:2008zc, ShuklaPhD,
    Margaryan:2018opu}. Crossed and permuted diagrams are not
  shown.}\label{fig:kernels}
\label{fig:transition}       
\end{figure}

\absatz
These results are obtained using the \emph{Transition-Density Method}
introduced in ref.~\cite{Griesshammer:2020ufp}.  It factorises the interaction
of a probe with a nucleus of $A$ nucleons into an \emph{interaction kernel}
between the probe and the $n$ \emph{active nucleons} which directly interact
with it, and a backdrop of $A-n$ \emph{spectator nucleons} which do not. The
effect of the latter is described by a \emph{$n$-body density}, namely a
transition probability density amplitude that $n$ active nucleons with a
specific set of quantum numbers are found inside the nucleus before an
interaction which transfers momentum $\qv$ and changes the initial quantum
numbers. This re-arranges the active nucleons into another specific set of
quantum numbers after the interaction. Figure~\ref{fig:transition} (top) illustrates
this separation, with the interaction kernel depicted as an arrow.  The one-
and two-body densities generated from a number of chiral potentials as well as
the AV18$+$UIX potential involving the \threeHe, \threeH and \fourHe system
are available at \url{https://datapub.fz-juelich.de/anogga}.

But what about interactions with more than $2$ active nucleons? It is a
fundamental advantage of \ChiEFT that it provides a well-defined procedure to
predict a hierarchy of $n$-body interaction kernels. As discussed in
refs.~\cite{Beane:1999uq, Beane:2004ra,Hildebrandt:2005ix, Hildebrandt:2005iw,
  Margaryan:2018opu} and summarised in~\cite[sect.~5.2]{Griesshammer:2012we},
only kernels with one and two active nucleons contribute in Compton scattering
up to and including $\calO(e^2\delta^4)$ [\NXLO{4}] at
$\omega\sim\mpi$. Therefore, three-or-more-body densities do not need to be
considered in the present $\calO(e^2\delta^3)$ [\NXLO{3}] investigation.

Another advantage of this factorisation is that densities of the same nucleus
and momentum transfer $\qv$ can be recycled for different interaction kernels,
while the same interaction kernels can be recycled in different nuclei. In
Compton scattering on \fourHe, we use the same kernels of
fig.~\ref{fig:kernels} as for our \threeHe results~\cite{Choudhury:2007bh,
  Shukla:2008zc, ShuklaPhD, Margaryan:2018opu} which originate in the deuteron
and proton/neutron kernels~\cite{Beane:1999uq, Beane:2004ra,
  Hildebrandt:2005iw, Hildebrandt:2005ix}. Thus, the few-body Compton
calculations share a common analysis framework. We also recycle the same
\threeHe and \fourHe densities in (for example) pion scattering and neutral
pion production~\cite{Long}. This split reduces the computational effort by
orders of magnitude: Densities are produced using well-developed modern
numerical few-body techniques and only once, while the kernel convolutions
only involve sums and integrals over the undetected one- and two-nucleon
quantum numbers and momenta. The summation over one-body quantum numbers is
near-instantaneous. For two-body matrix elements, 
integration over the relative momenta $\pv_{12}$ and $\pv^\prime_{12}$ and sum
over quantum numbers amount to less than a CPU hour per energy and angle on a
(s)lowly desktop for better-than-$0.7\%$ numerical accuracy. We reproduced
the ``traditional'' results 
for \threeHe~\cite{Choudhury:2007bh, Shukla:2008zc, ShuklaPhD,
  Margaryan:2018opu}, and the reduced computational cost allowed for more
extensive explorations of numerical convergence~\cite{Griesshammer:2020ufp}.

Let us turn now to the Compton kernel itself~\cite{Choudhury:2007bh,
  Shukla:2008zc, ShuklaPhD, Margaryan:2018opu}; see fig.~\ref{fig:kernels} (bottom). Its
one-nucleon part consists of the contributions in which photons interact with
a point nucleon (``Born terms'': top row on left), and of the ``structure
terms'' (other rows on left) which give rise to nonzero polarisabilities
because the photons couple to the $\Delta(1232)$ resonance and the pion cloud
around the nucleon and $\Delta$.  The two-nucleon part  (on the
right) describes the coupling of photons to charged meson-exchange
currents. Both kernels are complete up to and including order
$\calO(e^2\delta^3)$ [\NXLO{3}], where the small expansion parameter is
$\delta=\sqrt{\mpi/\Lambda_\chi}\approx\Delta/\Lambda_\chi\approx0.4$ with
$\Delta\approx300\;\MeV$ the $\Delta$-N mass splitting and
$\Lambda_\chi\approx700\;\MeV$ the breakdown scale of \ChiEFT.

\begin{figure}[!t]
\centering
\includegraphics[width=\linewidth,clip]
{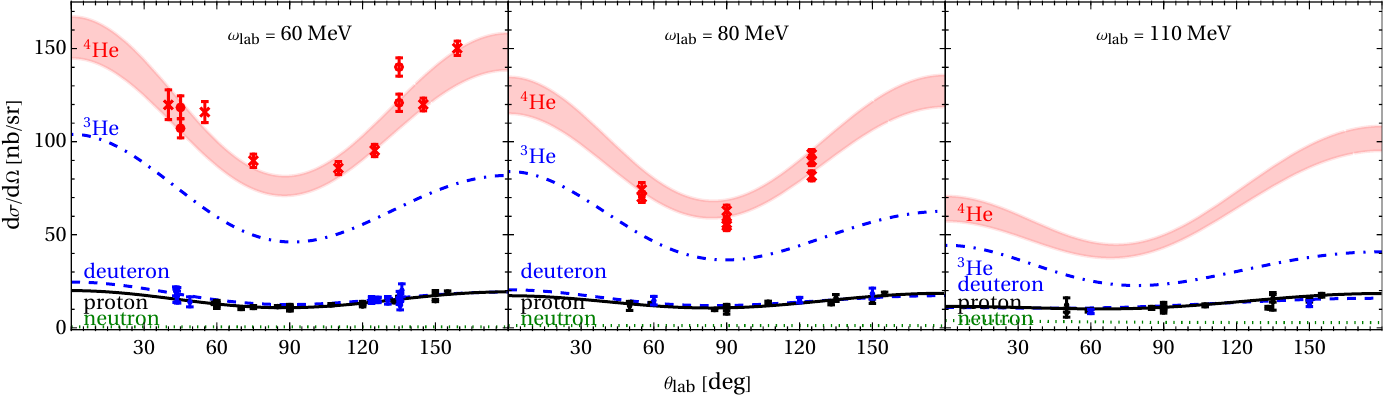}
\caption{(Colour on-line) \ChiEFT predictions and data at
  $\omegalab=\{60;80;110\}\;\MeV$ for a proton (black), neutron (green
  dotted), deuteron (blue dashed), \threeHe (blue dot-dashed) and \fourHe
  (band) target~\cite{ourstuff}.}
\label{fig:compare}       
\end{figure}

As fig.~\ref{fig:compare} shows, Compton cross sections on few-nucleon targets
do not scale with powers of the target charge $Z$ in the region of interest
for extracting polarisabilities, $\omega\in[50;120]\;\MeV$. While the proton
and deuteron cross sections ($Z=1$) are roughly of similar size, the \threeHe
and \fourHe differ by a factor of about two although they are both $Z=2$
targets. At $60\;(120)\;\MeV$, the average \fourHe nearly $7$ ($5$) times that
of the proton and deuteron. Rather, it appears that scaling is related to the
number of charged-pion currents (\ie~of pn pairs).

\absatz An upcoming publication will provide details of formalism,
uncertainties and results, sensitivity studies of the photon-beam asymmetry of
\fourHe, and a comparison to Compton scattering off lighter nuclear
targets~\cite{ourstuff}. \fourHe is an excellent choice because of the
expected high rates and clean signal. Work is under way to improve the theory
side to allow the step from predicting sensitivities which are adequate for
experimental planning, to extracting polarisabilities from high-accuracy
data. Indeed, data taking is scheduled at \HIGS in the very near future. The
theory group is also applying the transition-density formalism to Compton
scattering on other targets as well as to other processes on \threeHe,
\threeH, \fourHe~-- and beyond~\cite{Long}.



%
%
%

\subsection*{Acknowledgements}
M.~Ahmed, E.~Downie, G.~Feldman and M.~Sikora
provided feedback and encouragement. G.~Feldman and M.~Sikora tracked down
Compton data.
HWG thanks the organisers and participants of MENU 2023 in Mainz
for stimulating discussions and a delightful atmosphere.
This work was supported in part by the US Department of Energy under contracts
DE-SC0015393 (HWG, JL) and DE-FG02-93ER-40756 (DRP), by the UK Science and
Technology Facilities Council grant ST/P004423/1 (JMcG), and by the Deutsche
Forschungsgemeinschaft and the Chinese National Natural Science Foundation
through funds provided to the Sino-German CRC 110 ``Symmetries and the
Emergence of Structure in QCD'' (AN; DFG TRR~110; NSFC
11621131001). Additional funds for HWG were provided by \HIGS in concert with
Duke University, and by George Washington University. HWG's research was
conducted in part in GW's Campus in the Closet.
The nuclear densities were computed on \textsc{Jureca} and the
\textsc{Jureca-Booster} of the J\"ulich Supercomputing Centre (J\"ulich, Germany).


\end{document}